\documentclass[pre,twocolumn,aps,eqsecnum]{revtex4}
\usepackage{epsfig}

\begin{document}
\begin{titlepage}

\title{Fracture Toughness of Crystalline Solids}
\author{J.S. Langer}
\affiliation{Kavli Institute for Theoretical Physics, Kohn Hall, 
University of California, Santa Barbara, CA 93106}
\date{\today}
\begin{abstract}
This paper describes an attempt to construct a first-principles theory of the fracture toughness of crystalline solids.  It is based on the thermodynamic dislocation theory (TDT), which starts with the assertion that dislocations in solids must obey the second law of thermodynamics.  A second starting assumption is that fracture is initiated when the tip of a notch is driven to undergo a sharpening instability.  The results of this analysis are developed in comparison with measurements by Gumbsch and colleagues of the notch toughness of both predeformed and non-predeformed tungsten crystals.  The theory includes a mathematical conjecture regarding tip dynamics at  small dislocation densities.  Nevertheless, its predictions agree quantitatively with the experimental data, including both brittle and ductile fracture, over a wide range of temperatures, loading rates, and initial conditions.  
\end{abstract}
\maketitle

\end{titlepage}

\section{Introduction}
The strength of crystalline solids is one of the most important problems in materials theory; yet it has largely been ignored by theorists for about half a century because it has been thought to be unsolvable.  Of course, there is a large literature on this subject, exploring the dynamics of dislocations interacting with their environments.  But almost all of this literature is phenomenological; that is, it is not based on the fundamental physics of nonequilibrium processes. As a result, there are central  questions that remain unanswered. Remarkably, these questions are seldom even asked.   

For example, we have long known from observation that solids are stronger when they are colder; their yield stresses grow with decreasing temperature.  We also know that they become more brittle, i.e. they break more easily at lower temperatures despite the fact that they are stronger. How can these two properties be consistent with each other?  

It is only recently that we have begun to understand strain hardening and yield stresses for spatially uniform crystalline solids.  Cottrell and others since the 1950’s \cite{COTTRELL-53, HIRTH-LOTHE-68, COTTRELL-02} have asserted that this problem is impossibly difficult because the second law of thermodynamics –- according to them –- did not apply to dislocations.  They were wrong.  The second law says simply that driven complex systems always move toward states of higher probability, i.e. entropy must increase.  By ignoring this basic principle, and the fact that it had to be relevant to large systems of driven dislocations, Cottrell et al. provoked decades of curve-fitting and fundamentally incorrect phenomenologies.  

The thermodynamic dislocation theory (TDT) \cite{LBL-10,JSL-17rev,JSL-PCH-19} uses the second law to define an effective disorder temperature that captures the statistical features of crystalline systems in nonequilibrium situations.  With it, plus a simple model of thermally activated depinning of entangled dislocations, we now understand strain hardening and yield stresses in an oversimplified but basically realistic and experimentally tested \cite{JSL-LE-PNAS-21,LE-TWO LAWS-20} picture of crystalline solids.  However, the TDT by itself does not explain fracture toughness.

An explanation of brittle and ductile fracture requires a different kind of analysis.  Strain hardening and yielding can be spatially uniform phenomena; but the onset of  fracture, say, at the tip of a notch that is opening under stress, is a localized instability.  This is the lesson that we have learned recently from observations and theoretical analyses of brittle-ductile transitions in metallic glasses. \cite{RB-12,SCHetal-18,JSLBMG-20} There, the growing stress concentration at the tip of a notch generates a localized density of flow defects -- in that case shear transformation zones (STZ's) -- that controls the way in which the tip sharpens and either launches a propagating brittle crack or initiates large-scale ductile failure.  

My purpose here is to use the TDT in a similar kind of fracture analysis for crystalline materials. The two situations, while  different in major respects, have a great deal in common.  In both cases, the externally applied stress is strongly concentrated at a notch tip.  If this driving stress is applied rapidly enough, or if the plastic deformation is sufficiently slow or weak, then the concentrated elastic stress near the tip quickly reaches some critical value and a propagating crack is launched.  This is brittle behavior.  On the other hand, if plastic deformation is strong enough, the notch tip becomes shielded by a plastic boundary layer that suppresses elasticity-induced fracture.  Then, when the far-field stress becomes large enough, this boundary layer expands rapidly and the tip loses its shielding.  At this point, the tip stress grows suddenly, thus initiating some kind of large-scale ductile failure whose details -- like the details of brittle crack propagation -- vary from system to system and are beyond the scope of this analysis.  

In the absence of a predictive theory of crack initiation, it is not surprising that there exists only little experimental data relevant to the picture that I will develop here.  Most of the literature focuses on the behaviors of single dislocations, or small groups of them, in the neigborhoods of mathematically sharp crack tips or other kinds of heterogeneities, and thus does not probe the larger-scale dynamics that I assert is more important. Notable examples of earlier research of this kind include \cite{ARGON-01,TANAKAetal-08,ASTetal-18}.

So far as I know, the most relevant experimental observations are those of Gumbsch and coworkers.\cite{GUMBSCHetal-98,GUMBSCH-03} I find their results to be very valuable because they confirm central aspects of the physics-based TDT approach to fracture dynamics and also point to places where that theory is seriously incomplete.  Gumbsch {\it et al} studied temperature-dependent notch toughness in tungsten, in both undeformed single crystals and predeformed (i.e. ``prehardened'') samples.  The crucial difference between the two kinds of measurements was that the predeformed systems had absorbed some energy of work hardening and thus, initially, contained substantial densities of dislocations.  This is a simpler situation than that of the undeformed crystals because we can start by assuming that the dislocations are dense enough near the notch tip that the TDT entanglement theory is valid. 

By focussing first on this case, we can see how plastic shielding at the notch tip is governed by the strongly stress and temperature dependent TDT strain rate, which determines a sharply defined yield stress and a plastic boundary layer.  The onset of ductile failure is triggered by a nonlinear heating instability analogous to that seen in adiabatic shear banding.\cite{MARCHAND-DUFFY-88}  This analysis also allows us to determine values of some parameters that are needed for studying the undeformed single-crystal experiments. 

The second part of this paper is devoted to Gumbsch's  non-predeformed crystals, where the initial density of dislocations must be so small that the average spacing between them is greater than the radius of the notch tip.  In this case, I propose that early-stage plasticity is determined by strongly temperature dependent dislocation drag forces.  As work is done on this system by the notch-opening stress, dislocations are created and move out independently according to the drag dynamics. At low enough temperatures, where the drag force is small, the resulting plastic deformation sharpens the tip and induces brittle fracture.  At higher temperatures, the drag forces become too large, the dislocations move too slowly, and there is a transition from independently moving to strongly entangled dislocations. This is the theoretically most difficult and speculative part of my analysis.  But it is potentially the most important part because it raises new fundamental questions. 

\section{Elliptical Approximation}
\label{Ellipse}

The starting point for this analysis, as in \cite{JSLBMG-20}, is an elliptical approximation for the time dependent shape of the notch at which fracture is initiated. 

Consider an incompressible plate of elasto-plastic material lying in the  $x,y$ plane and containing an elliptical hole.  The ellipse is elongated in the $x$ direction; and a mode-I stress $\sigma_{\infty}$ is imposed in the $y$ direction very far from the hole.  Assume hypo-elasto-plasticity (additive decomposition of elastic and plastic rates of deformation). 

The static, linearly elastic version of this problem has been solved by Muskhelishvili \cite{MUSK-63}.  The first step is to transform from Cartesian coordinates $(x,y)$ to elliptical coordinates ($\zeta$, $\theta$):
\begin{equation}
\label{zetaW}
x=W\left(\zeta+{m\over\zeta}\right)\,\cos\,\theta, ~~~y=W\left(\zeta - {m\over\zeta}\right)\,\sin\,\theta.
\end{equation}
Curves of constant $\zeta$ are ellipses, and curves of constant $\theta$ are orthogonal hyperbolas. If we take the boundary of the elliptical hole to be at $\zeta=1$, then the semi-major and semi-minor axes of the ellipse have lengths $W(1+m)$ and $W(1-m)$ respectively. Let $0<m<1$ so that the long axis of the ellipse lies in the $x$ direction, perpendicular to the applied stress, in analogy to a mode-I crack. 

To produce a long, thin ellipse, let $W$ be larger than any other length in the system, and fix $m\le 1$ so that the curvature at the tip, i.e. at $x=W(1+m)$, is large but finite.  Denote this curvature by ${\cal K}_{tip}$. Then a calculation to leading order in $1/\sqrt{W}$ yields
\begin{equation}
\label{m}
m \approx 1-2\,\epsilon;~~\epsilon\equiv\sqrt{1\over 2 \,{\cal K}_{tip} W}\ll 1,
\end{equation}
where $\epsilon$ will be the principal small parameter in this analysis.

My scheme is to use the elasto-plastic equations of motion to determine the behavior of this elliptical notch under  mode-I straining.  There are serious difficulties with this scheme.  We know that this shape does not remain elliptical; its motion must involve shape changes that cannot be described simply by time dependent values of the  parameters $W$ and  ${\cal K}_{tip}$.  This is true even in the purely elastic case for a time-varying applied stress.   

To minimize the difficulties of the elliptical approximation, we can focus on only the immediate neighborhood of the sharp tip, $\theta < \epsilon \ll 1$, and look only at the early onset of plastic deformation there.  For simplicity, I have relegated the general elliptical formulas to the Appendix, and use only near-tip and early-onset approximations derived from those formulas here in the main text.  

Equation (\ref {sigmaeqn2}) tells us that the deviatoric stress near the tip is
\begin{equation}
\label{stress1}
s_{\theta,\theta}= - s_{\zeta,\zeta} \equiv s(\tilde x,\theta) \approx {\sigma_{\infty}\,\epsilon^2\over (\epsilon + \tilde x)^3}\,\Bigl(1- {2\,\theta^2\over \epsilon^2}\Bigr),
\end{equation}
where $\tilde x = \zeta - 1 \ll 1$ is the dimensionless distance from the tip along the $x$ axis.  The bare tip stress is
\begin{equation}
s(0,0) = {\sigma_{\infty}\over \epsilon} = \sigma_{\infty}\,\sqrt{2 \,\kappa \,W/d_{tip}}
\end{equation}
where $ \kappa \equiv {\cal K}_{tip}\,d_{tip}$ is a dimensionless tip curvature, $d_{tip}$ is the initial tip radius, and $\mu$ is the shear modulus.  The factor $(W/d_{tip})^{1/2}$ is a measure of the initial stress concentration at the tip.  

For simplicity, assume that the material is incompressible.  Also assume hypo-elasto-plasticity (additive decomposition of elastic and plastic rates of deformation). These assumptions imply that the diagonal elements of the rate-of-deformation tensor have the form
\begin{eqnarray}
\label{RODT}
\nonumber 
 && D_{\theta\theta}(\zeta,\theta) = - D_{\zeta\zeta}(\zeta,\theta)\cr\\&&\equiv D(\zeta,\theta)\cong {1\over 2 \mu}{ds(\tilde x,\theta)\over dt}+ D^{pl}(\tilde x,\theta),
 \end{eqnarray}
 where $s(\tilde x,\theta)$ is the deviatoric stress defined above.  $D^{pl}(\tilde x,\theta)$ is the ($\theta$,$\theta$) element of the plastic rate-of-deformation tensor.  It will play a central role in what follows.  

Next use Eqs.(\ref{Dzeta1}) and (\ref{Dtheta1}) in the Appendix to express the rate of deformation tensor ${\cal D}$ in terms of the material velocities  $v_{\zeta}$ and $v_{\theta}$ near the crack tip, and thus use Eq.(\ref{RODT}) to write equations of motion for those velocities.  Using the same approximations for small $\tilde x$ and small $\theta$ used above, I find
\begin{eqnarray}
\label{Dzetazeta2}
\nonumber
 {\cal D}_{\zeta\zeta}&&\cong {1\over 2 \epsilon W}\,\Bigl[{\partial v_{\zeta}\over \partial\tilde x}+ {\partial v_{\theta}(0)\over\partial \theta}\,{\theta^2\over \epsilon^2}\Bigr]\,\Bigl(1 - {\theta^2\over 2\,\epsilon^2}\Bigr)\cr\\&& \equiv -\, D(\tilde x,\theta);
\end{eqnarray}
and
\begin{eqnarray}
\label{Dthetatheta2}
\nonumber
{\cal D}_{\theta\theta}
&&\cong {1\over 2 \epsilon W}\,\Bigl[{\partial v_{\theta}\over \partial\theta}+ {v_{\zeta}\over\epsilon}\,\Bigl(1- {\theta^2\over \epsilon^2}\Bigr)\,\Bigr]\Bigl(1 - {\theta^2\over 2\,\epsilon^2}\Bigr)\cr\\&& \equiv + \,D(\tilde x,\theta),
\end{eqnarray}
where 
\begin{equation}
\label{Dxtheta}
D(\tilde x,\theta) \cong D_0(\tilde x) \Bigl(1- {2 \,\theta^2\over \epsilon^2}\Bigr),
\end{equation}
and
\begin{equation}
\label{D0}
D_0(\tilde x) = {1\over 2\mu} \dot s(\tilde x,0)  + D_0^{pl}(\tilde x).
\end{equation}
In Eqs.(\ref{Dxtheta}) and (\ref{D0}), I have assumed that, to lowest order in $\theta^2/\epsilon^2$, $D_0^{pl}$ carries the same $\theta$ dependence as the stress.

The tip velocity is 
\begin{equation}
\label{vtip}
v_{tip} = v_{\zeta}(0)=-\int_0^{\infty} d\tilde x\,{dv_{\zeta}\over d\tilde x}= 2\epsilon W \int_0^{\infty} d\tilde x\, D_0(\tilde x).~~~~~
\end{equation}
We also need to compute the tip curvature.
Start with the geometric formula \cite{JSL-87}
\begin{equation}
\label{Ktip}
-{\dot{\cal K}_{tip}\over {\cal K}^2_{tip}} =  v_{tip} + {1\over 2\,{\cal K}_{tip}\,W}\,{\partial^2 v_{\zeta}\over \partial\,\theta^2}\Bigr|_{\theta = 0}.
\end{equation}
To evaluate this expression, define 
\begin{equation}
v_{\zeta}(\tilde x,\theta) \equiv v_0(\tilde x)+v_2(\tilde x)\, {\theta^2\over\epsilon^2},
\end{equation}
so that Eq.(\ref{Ktip})  becomes
\begin{equation}
- {\dot \kappa\over \kappa} = {\kappa\over d_{tip}}\,\bigl(v_0(0) + 2\,v_2(0)\bigr).
\end{equation}
Now use Eq.(\ref{Dthetatheta2}) at $\theta=0$ to write 
\begin{equation}
\Bigl({\partial v_{\theta}\over \partial\theta}\Bigr)_{\theta=0} =- {v_0(\tilde x)\over\epsilon} + 2\,\epsilon\,W D_0(\tilde x),
\end{equation}
and insert this into Eq.(\ref{Dzetazeta2}).  Collecting terms proportional to $\theta^2/\epsilon^2$, I find 
\begin{equation}
{d v_2\over d \tilde x} = \epsilon\,W\,D_0(\tilde x) +{v_0(\tilde x)\over\epsilon}.
\end{equation}
Then, using
\begin{equation}
\label{v0eqn}
{d v_0\over d \tilde x} = -2\epsilon\,W\,D_0(\tilde x)
\end{equation}
and combining terms, I find
\begin{equation}
\label{kappadot}
{\dot \kappa\over \kappa}={2\over \epsilon^2}\int_0^{\infty} \tilde x d\tilde x \,D_0(\tilde x).
\end{equation}

\section{TDT Plasticity and the boundary layer approximation at large dislocation densities}

One of the most important features of the thermodynamic dislocation theory is its first-principles prediction of yield stresses.  As shown, e.g. in \cite{JSL-17rev}, (and as will be shown again here), the TDT deformation mode changes very sharply but continuously from reversible elasticity to irreversible plasticity at a yield stress $s_y$ that depends on  temperature, dislocation density, and strain rate.  Understanding this transition is especially important for fracture analysis.  It tells us that, when the externally applied stress becomes large enough, the deformation changes abruptly from elastic to plastic at some position in front of the notch tip, thereby producing a plastic zone whose behavior controls the onset of fracture.  Knowing the dynamics of this elastic-plastic boundary allows us to formulate a boundary-layer approximation similar to the one that has been used in the metallic-glass theory.\cite{JSLBMG-20}

The TDT expression for the plastic rate of deformation along the $\tilde x$ axis, $D_0^{pl}(\tilde x)$, first appearing above in Eq.(\ref{D0}), is
\begin{equation}
\label{Dpl0}
D_0^{pl}(\tilde x) = {1\over \tau_{pl}}\sqrt{\tilde\rho}\, \exp\Bigl[ - {T_P\over T}\,e^{- \tilde s(\tilde x)/\sqrt {\tilde\rho}}\Bigr].
\end{equation}
Here, the plastic time scale is $\tau_{pl} = a\,\tau_0/b$, where $a$ is a minimum spacing between dislocations, $b$ is the length of the Burgers vector, and $\tau_0$ is a microscopic time scale of the order of $10^{-10}\,s$.  $\tilde\rho$ is the density of dislocations in units of $a^{-2}$; $\tilde s = s/\mu_T$, where $\mu_T$ is a reduced shear modulus, and $\mu_T\,\sqrt{\tilde\rho}$ is the Taylor stress (see \cite{JSL-17rev, JSL-PCH-19}).  This double-exponential formula assumes  that plastic deformation is entirely determined by the rate at which entangled dislocations are unpinned from each other by thermal fluctuations.  The pinning energy is $k_B T_P$. Note that $D_0^{pl}$ is an extremely rapidly varying function of $\tilde s$,  $\tilde\rho$, and the ambient temperature $T$ because the pinning energy is very large, $T_P \sim 10^4\,K$.  

Equation (\ref{D0}) gives the total rate of deformation $\cal{D}_{\theta\theta}$ along the $\tilde x$ axis as the sum of elastic and plastic terms.  Yielding occurs when the elastic term $\dot s/2\,\mu$ becomes vanishingly small because it is dominated by the much more strongly stress dependent plastic term.  That is, the yield point occurs when
\begin{equation}
{1\over \tau_{pl}}\sqrt{\tilde\rho} \exp\Bigl[ - {T_P\over T}\,e^{- \tilde s_y/\sqrt {\tilde\rho}}\Bigr] = D_y,
\end{equation}
where $D_y$ is the total rate of deformation at the yield point and $\tilde s_y$ is the dimensionless yield stress in units of $\mu_T$. Thus
\begin{equation}
\label{sy1}
{\tilde s_y\over \sqrt{\tilde\rho}} = \ln\Bigl({T_P\over T}\Bigr) - \ln\Bigl(\ln{\sqrt{\tilde\rho}\over \tau_{pl} D_y}\Bigr).
\end{equation}

To estimate $D_y$, note that this total deformation rate must be continuous across the elastic-plastic boundary, and look at the outer edge of this boundary where the elastic stress is given by Eq.(\ref{stress1}).  Also note that we need only an order-of-magnitude estimate for $D_y$  because it appears here only as the argument of a double logarithm. Write Eq.(\ref{stress1}) in the form:
\begin{equation}
\label{stress2}
s(\tilde x_y,0) = s_y = {\sigma_{\infty}\,\epsilon^2\over (\epsilon + \tilde x_y)^3} \equiv {\sigma_{\infty}\over \epsilon \,\nu^3},
\end{equation} 
where $\tilde x_y = \epsilon (\nu -1)$ is the position of the  boundary, and 
\begin{equation}
\label{nudef}
\nu^3 \equiv {\sigma_{\infty}\over s_y \epsilon}
\end{equation}

To find $D_y$, take the time derivative of Eq.(\ref{stress2}):
\begin{equation}
D_y = {1\over 2\mu}\,\dot s(\tilde x_y,0) \cong {\dot\sigma_{\infty}\over 2\mu\epsilon \,\nu^3} = c_0 {\dot \sigma_{\infty}\over \sigma_{\infty}}\tilde s_y,
\end{equation}
where $c_0 = \mu_T/2\,\mu$.  Here, I have assumed that $\sigma_{\infty}$ carries the dominant time dependence and that $\nu$ and $\epsilon$, which describe the position and curvature of the tip and yield surface, vary more slowly.  Also note that the ratio $\dot \sigma_{\infty}/ \sigma_{\infty}$ is an inverse time scale that, for these purposes, we can identify as the external driving rate $\tau_{ex}^{-1}$. (I will be more careful later in defining $\tau_{ex}$.)  Combine this result with Eq.(\ref{sy1})  to obtain
\begin{equation}
\label{sy2}
{\tilde s_y\over \sqrt{\tilde\rho}} = \ln\Bigl({T_P\over T}\Bigr) - \ln\Bigl(\ln{\xi\sqrt{\tilde\rho}\over c_0\, \tilde s_y}\Bigr),
\end{equation}
where, as in \cite{JSLBMG-20}, I have defined $\xi \equiv \tau_{ex}/\tau_{pl}$.  Eq.(\ref{sy2}) is a nonlinear equation that can be solved for $\tilde s_y/ \sqrt{\tilde\rho}$; but it is sufficient to make a first-order approximation for large $T_P/T$:
\begin{equation}
\label{sy3}
{\tilde s_y\over \sqrt{\tilde\rho}} \cong \ln\Bigl({T_P\over T}\Bigr) - \ln\Bigl[\ln\Bigl({10\,\xi\over c_0\,\ln(T_P/T)}\Bigr)\Bigr].
\end{equation}
Here, the arbitrary factor $10$ in the argument of the  double logarithm simply keeps it greater than unity for computational purposes. 

In order for $\tilde s_y$ to be well defined, we must constrain this formula to be valid only for large enough $\tilde\rho$. Otherwise, small values of $\tilde\rho$ would produce small values of $\tilde s_y$ and large values of $\nu$ according to Eq.(\ref{nudef}); and this would mean that the width of the plastic zone, $\tilde x_y =\epsilon\,(\nu -1)$, would be unphysically large. Moreover, we know that this width cannot be negative; $\nu$ cannot be less than unity.  Thus, I rewrite Eq.(\ref{nudef}) as follows:
\begin{equation}
\label{nudef2}
\nu_1(y) \equiv \cases{y^{1/3},&if $y>1$,\cr ~~~1,&otherwise}
\end{equation}
and then
\begin{equation}
\label{nudef3}
\bar\nu(y) = \cases{\nu_1(y),&if  $\tilde\rho>\tilde{\rho}_{min},$\cr~~~ 1,&otherwise,}
\end{equation}
where where $y = \sigma_{\infty}/ s_y \epsilon$.  The quantity $\tilde{\rho}_{min}$ is a minimum dislocation density required for the validity of this statistical theory in the neighborhood of a finite-sized notch tip. 

The next step toward a boundary-layer theory is an approximation for the integrations over the plastic zones in Eqs.(2.10) and (2.17).  I approximate $\tilde{s}(x)$ by writing
\begin{equation}
\label{BLA}
\tilde{s}(\tilde x)\cong \tilde{s}_y+ (\tilde{s}_0 - \tilde{s}_y)\Bigl(1 - {\tilde x\over \tilde {x}_y}\Bigr)~~~{\rm for}\,0<\tilde {x}<\tilde {x}_y,
\end{equation}
where $\tilde s_0$ is the time-dependent tip stress.  The outer boundary, at $\tilde x= \tilde x_y $, is determined by the yield stress $\tilde s_y$ as computed above. In principle, we could substitute this approximation into the formula for $D_0^{pl}(\tilde x)$ given in Eq.(\ref{Dpl0}) and evaluate the integrals.  A  much simpler procedure is to write
\begin{equation}
\label{Dpl0q}
D_0^{pl}(\tilde x) \cong {1\over \tau_{pl}}\,q(\tilde s_0,\tilde\rho,T)\,\bigl(1-{\tilde x\over\tilde x_y}\bigr), ~~0 < \tilde x < \tilde x_y,
\end{equation}
where
\begin{equation}
\label{qdef}
q(\tilde s_0,\tilde\rho,T)= \sqrt{\tilde\rho}\, \exp\Bigl[ - {T_P\over T}\,e^{- \tilde s_0/\sqrt {\tilde\rho}}\Bigr].
 \end{equation}

Outside the plastic zone, i.e. for $\tilde x > \tilde x_y$, the stress is determined by Eq.(\ref{stress1}) at $\theta = 0$. Inserting the latter expression and the approximations in Eqs.(\ref{BLA}) and       (\ref{Dpl0q}) into Eqs.(\ref{vtip}) and (\ref{kappadot}), and integrating separately over the elastic and plastic zones, I find
\begin{equation}
\label{vtip2}
{v_{tip}\over d_{tip}} = {(\bar\nu -1)\over 2\kappa}  \Bigl( c_0 \dot {\tilde s}_0 + {q(\tilde s_0,\tilde\rho,T)\over \tau_{pl}} \Bigr)+ \Bigl({1\over 2\, \kappa\,\bar\nu^2}\Bigr) \Bigl({\dot\sigma_{\infty}\over 2\,\mu\,\epsilon }\Bigr)
\end{equation}
and
\begin{equation}
\label{kappadot2}
{\dot\kappa\over \kappa}=  {(\bar\nu -1)^2\over 3}\Bigl( c_0 \dot{\tilde s}_0 + {q(\tilde s_0,\tilde\rho,T)\over \tau_{pl}} \Bigr)+ \,\Bigl({2\bar\nu -1\over \bar\nu^2}\Bigr) \Bigl({\dot\sigma_{\infty}\over  2\,\mu\,\epsilon }\Bigr).
\end{equation}

It remains to find an equation of motion for the tip stress $\tilde s_0$.  I do this by making a circular approximation as in \cite{JSLBMG-20}.  Consider a pair of concentric circles around the notch tip, with radial variable $r$ and a radial rate of deformation $v(r)$. The inner circle has a radius $R$ equal to the tip radius $d_{tip}/\kappa$, and the outer circle is at the boundary of the plastic zone, thus at $R_1 = \bar\nu R$.  The analogs of the equations of motion, Eqs.(\ref{Dzetazeta2}) and (\ref{Dthetatheta2}), are 
\begin{equation}
\label{dvdr1} 
{\partial v\over \partial r} + {v\over r} = 0,
\end{equation}
and
\begin{equation}
\label{dvdr2}
- {\partial v\over \partial r} + {v\over r} =- r{d\over dr}\Bigl({v\over r}\Bigr)=2\,\Bigl(c_0\,\dot{\tilde s}(r) + D_0^{pl}(r)\Bigr). 
\end{equation}
The first of these equations is the statement of incompressibility, which implies that $v(r) = R\,\dot R/r$.  If we make the boundary-layer approximations analogous to Eqs.(\ref{BLA}) and (\ref{Dpl0q}), that is:
\begin{equation}
\label{BLA0}
\tilde s(r) \cong \tilde s_y + (\tilde s_0 - \tilde s_y)\,{R_1 - r\over R_1 - R},~~~R < r < R_1,
\end{equation}
and
\begin{equation}
\label{Dpl0qr}
D_0^{pl}(r) \cong {1\over \tau_{pl}}\,q(\tilde s_0,\tilde\rho,T)\,\Bigl( {R_1-r\over R_1-R}\Bigr), ~~R < r < R_1,
\end{equation}
then we can integrate Eq.(\ref{dvdr2}) and use Eq.(\ref{Dpl0qr}) to find
\begin{equation}
\label{Rdot}
{\dot R\over R} - {\dot R_1\over R_1} = \Bigl(c_0 \dot {\tilde s}_0 + {q(\tilde s_0,\tilde\rho,T)\over \tau_{pl}}\Bigr)\,\lambda(\bar\nu)
\end{equation}
where, using $R_1/R = \bar\nu$,
\begin{equation}
\lambda(\bar\nu) = 2\,\int_R^{R_1} {dr\over r}\Bigl({R_1 - r\over R_1-R}\Bigr)={2 \,\bar\nu\over \bar\nu-1}\,\ln \bar\nu - 2.
\end{equation}
Finally, use Eq.(\ref{vtip2}) to evaluate $\dot R= v_{tip}$, integrate Eq.(\ref{v0eqn}) to evaluate $\dot R_1$, insert these expressions into the left-hand side of Eq.(\ref{Rdot}), and solve for $\dot{\tilde  s_0}$.  The resulting equation of motion for the tip stress is
\begin{equation}
\label{s0dot}
c_0 \dot {\tilde s}_0 = - {q(\tilde s_0,\tilde\rho,T)\over \tau_{pl}} + \Bigl({\dot\sigma_{\infty}\over 2\,\mu\epsilon }\Bigr)\,{\Lambda(\bar\nu)\over \bar\nu^3},
\end{equation}
where 
\begin{equation}
\label{Lambdadef}
\Lambda(\bar\nu)= {\bar\nu -1\over 2\,\lambda(\bar\nu) - \bar\nu +1} = {(\bar\nu -1)^2\over 4\,\bar\nu \,\ln \bar\nu - (3 + \bar\nu)(\bar\nu -1)}.~~~~~~
\end{equation} 
Despite appearances, $\Lambda(\bar\nu)$ is continuous at the onset of plasticity, i.e. $\Lambda(1) =1$.  Importantly, it diverges at $\bar\nu \cong 5.1$, describing the sudden expansion of the plastic zone and rapid unshielding of the notch tip that occurs when the far-field stress becomes critically large.  

The next step in this analysis is to rewrite the preceding equations of motion in a dimensionless notation and to supplement them by equations of motion for the internal variables $\tilde\rho$ and $T$.  First, define a dimensionless stress intensity:  
\begin{equation}
\label{psidef}
\psi \equiv {\sigma_{\infty}\over \mu_T} \sqrt{2\,W\over d_{tip}} = {\sigma_{\infty} \over \mu_T\,\epsilon\,\sqrt{\kappa}};
\end{equation}
and define
\begin{equation}
\label{dotpsidef}
\label{dotpsi}
\dot\psi = {\dot\sigma_{\infty}\over \mu_T} \sqrt{2\,W\over d_{tip}} \equiv {1\over \tau_{ex}};~~~\xi\equiv {\tau_{ex}\over \tau_{pl}}.
\end{equation}
Assuming that the external driving rate $\dot\sigma_{\infty}$ remains constant, we can use $\psi$ as the independent time-like variable, and rewrite Eqs. (\ref{kappadot2}) and (\ref{s0dot}): 
\begin{eqnarray}
\label{kappadot3}
{1\over \kappa^{3/2}}{d\kappa\over d\psi}&=&  c_0\Bigl[{(\bar\nu -1)^2\over 3}{\Lambda(\bar\nu)\over\bar\nu^3} + \Bigl({2 \bar\nu -1\over \bar\nu^2}\Bigr) \Bigr],~~~~
\end{eqnarray}
and
\begin{equation}
\label{tipstress}
{d \tilde s_0\over d\psi} = - {\xi\over c_0} q(\tilde s_0,\tilde\rho,T) + \sqrt{\kappa}\, {\Lambda(\bar\nu)\over\bar\nu^3}.
\end{equation}
Here, $\bar\nu$ remains as given by Eqs.(\ref{nudef2}) and (\ref{nudef3}) with $y = \psi \sqrt{\kappa}/ \tilde s_y$.  The yield stress $\tilde s_y$ is given by Eq.(\ref{sy3}) and the function $\Lambda(\bar\nu)$ is given by Eq.(\ref{Lambdadef}). The expression for $v_{tip}$ in Eq.(\ref{vtip2}) was used in the derivation of Eq.(\ref{s0dot}) and is no longer needed at this stage of the analysis.

We also need equations of motion for the dislocation density $\tilde\rho$ and the temperature $T$. In principle, these quantities should be spatially varying fields; but it will be consistent with the other approximations made here to treat them as single values at the notch tip.  Thus I write:
\begin{equation}
\label{rho-psi}
{d \tilde\rho\over d\psi} = A\,\xi \,q(\tilde s_0,\tilde\rho,T)\,\tilde s_0(\psi)\,\Bigl[1 - {\tilde\rho(\psi)\over \tilde\rho_{\infty}}\Bigr].
\end{equation}
This is a simplified version of the usual TDT equation that says that the rate at which dislocations are formed is proportional to the rate at which plastic work is done in units of the dislocation energy.  The right-hand side contains a detailed-balance factor, $[1 - \tilde\rho(\psi)/ \tilde{\rho}_{\infty}]$, which accounts for dislocation annihilation by requiring that $\tilde\rho$ approach its effective thermodynamic steady-state value $\tilde\rho_{\infty}$.  The dimensionless coefficient $A$ should be approximately temperature-independent if the dislocation energy is a constant.  As we shall see, however, $A$ may depend on the driving rate ({\it via} the dimensionless ratio $\xi$) because of geometric effects ignored in this analysis. 

\begin{figure}[h]
\begin{center}
\includegraphics[width=\linewidth] {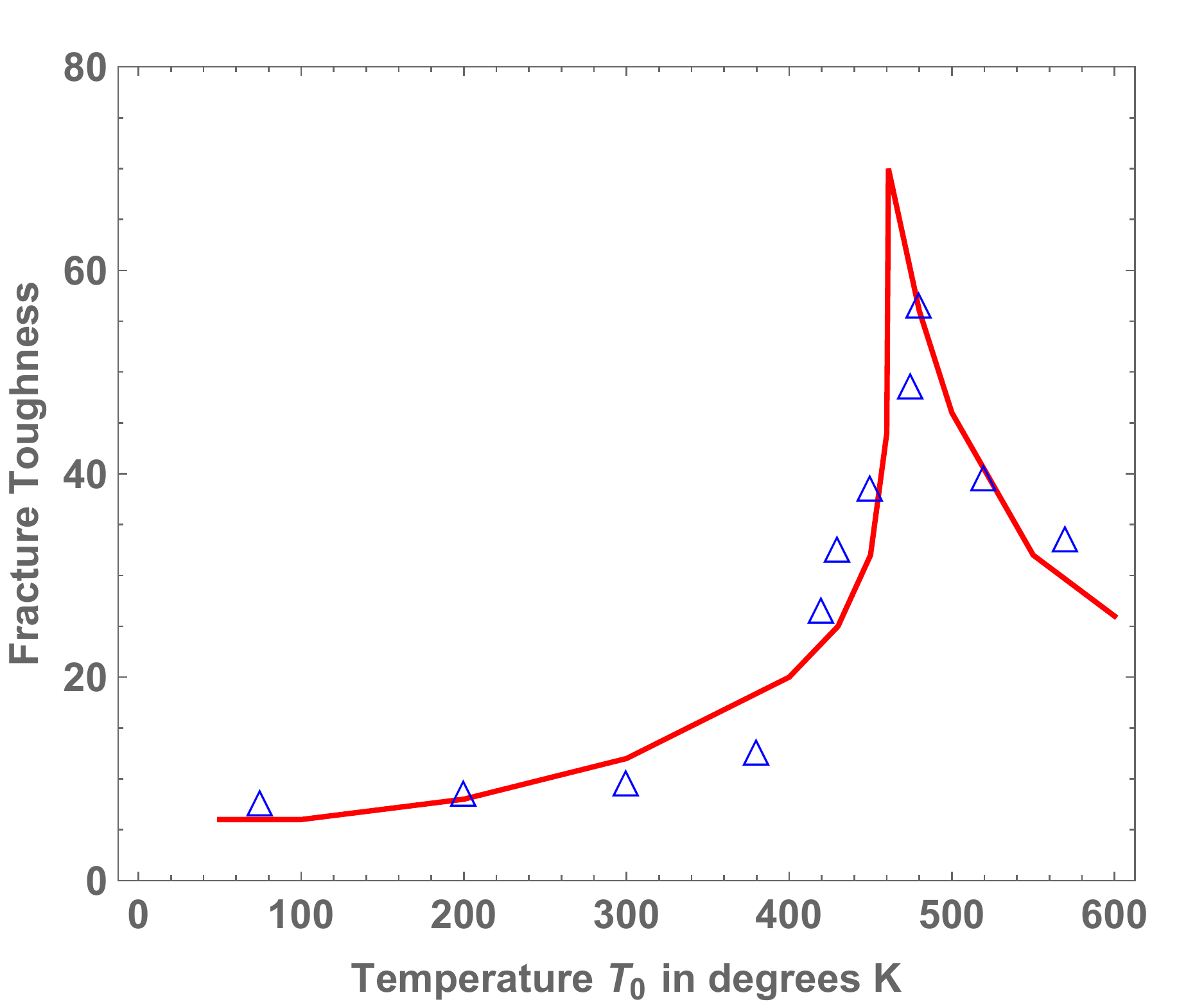}
\caption{Experimental data for fracture toughness as a function of temperature (open triangles), and theoretical prediction (solid curve), for predeformed tungsten. The fracture toughness  ${\cal K}_c$ is in units $M\!Pa\,\,m^{1/2}$. The loading rate, {\it i.e.} the stress-intensity rate, is $0.1 M\!Pa\,\,m^{1/2}\,s^{-1}$ .}   \label{BDGFig1}
 \end{center}
\end{figure}

Similarly, I assume an equation of motion for the tip temperature $T$ in the form:
\begin{equation} 
\label{TQ}
{d T\over d\psi} = C(T)\,\xi \,q(\tilde s_0,\tilde\rho,T)\,\tilde s_0(\psi) ,
\end{equation}
where $C(T)$ is proportional to the Taylor-Quinney conversion factor, {\it i.e.} the fraction of the work done on the system that is converted to heat.  In principle, there should be a cooling term in Eq.(\ref{TQ}) describing how $T$ relaxes to the ambient temperature $T_0$.  Here I will assume that that term is negligible.  However, as in earlier analyses of thermal softening and adiabatic shear banding \cite{JSL-17,LTL-17,LTL-18}, the coefficient $C(T)$ is strongly $T$-dependent.  

\section{Comparison with experiment at large dislocation densities}

Turn now to the use of the preceding equations in interpreting Gumbsch's data for fracture toughness of predeformed tungsten.  Both the experimental data and the theory are shown in Fig. 1.  The open triangles are the experimental points taken from Fig. 2 of \cite{GUMBSCH-03}; the solid curve is the theory. As stated in the Introduction, the principal difference between these predeformed systems and non-predeformed single crystals is that here we can assume that the initial density of dislocations is high enough to be in the TDT entanglement regime. 

\begin{figure}[h]
\begin{center}
\includegraphics[width=\linewidth] {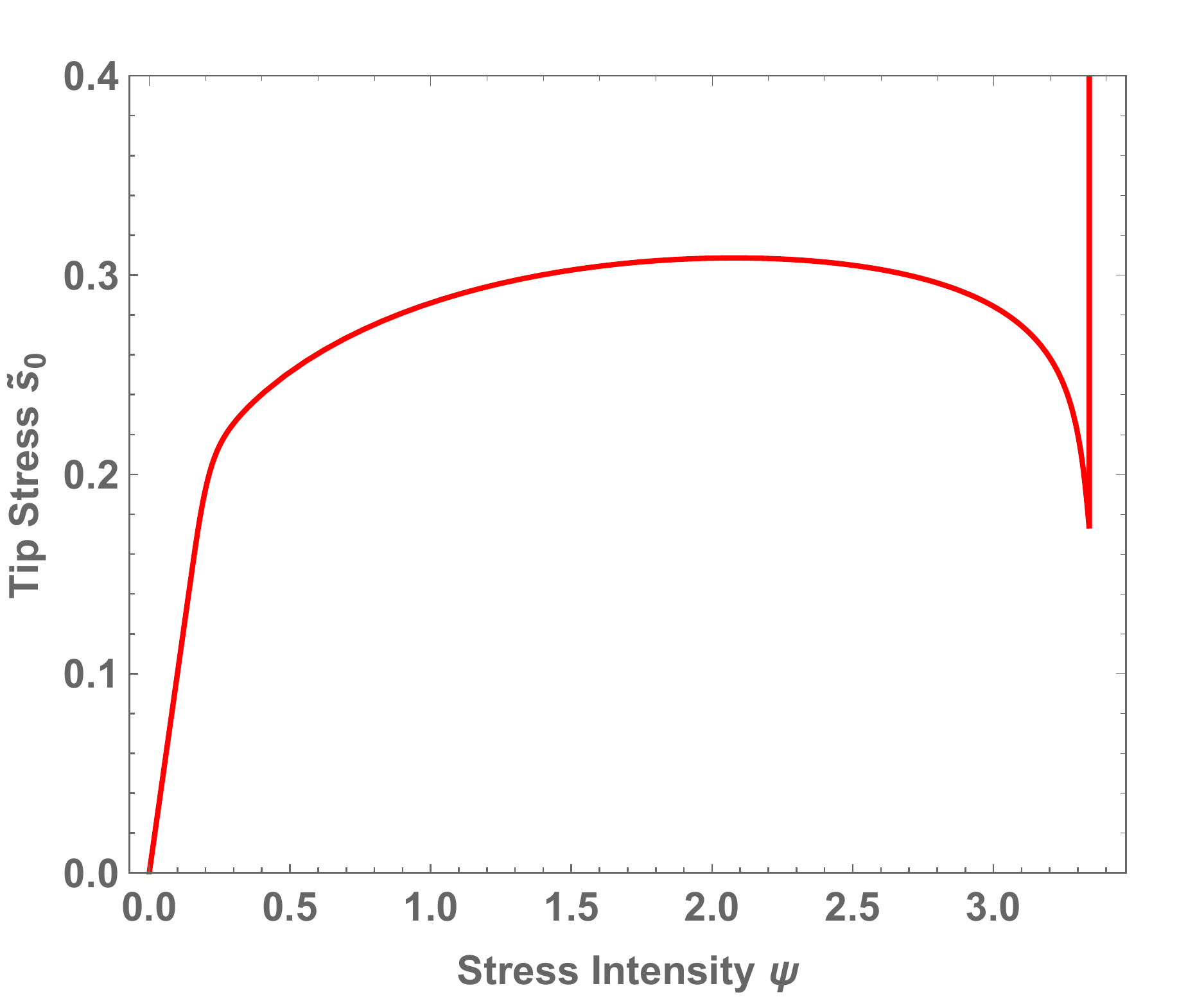}
\caption{Theoretical tip stress $\tilde s_0(\psi)$ for predeformed Tungsten at initial temperature $T_0 = 460\,K$. An enlarged graph of the singularity at $\psi \cong 3.34$ is shown in Fig \ref{BDGFig5}. }   \label{BDGFig2}
 \end{center}
\end{figure}

 {\it System Parameters}: Throughout this paper, I have used $\tilde\rho_{\infty} = e^{-4}$ \cite{LBL-10}, and have chosen $\tilde\rho_{min} = 0.1\,\tilde\rho_{\infty}$.  The initial dislocation density for these predeformed samples is taken to be $\tilde\rho_0 = 0.2\,\tilde\rho_{\infty}$. Le and collaborators \cite{KCL?} have found recently from bulk plasticity data that $T_P \cong 36,000\,K$ for tungsten.  In Eq.(\ref{rho-psi}), I have set $A= 10$. In Eq.(\ref{TQ}),  I have used 
\begin{equation}
\label{TQ2}
C(T) = C_0\,\exp(- T_A/T)
\end{equation}
 with $T_A = 3500\,K$ and $C_0 =0.7 \times 10^8$.  I have arbitrarily set $\xi = 1$ for this first set of calculations, which means that I have absorbed an arbitrary dimensionless factor into the ratio of time scales $\tau_{ex}/\tau_{pl}$. Throughout this analysis, I have used $c_0 \equiv \mu_T/2\,\mu = 0.01$.     

\begin{figure}[h]
\begin{center}
\includegraphics[width=\linewidth] {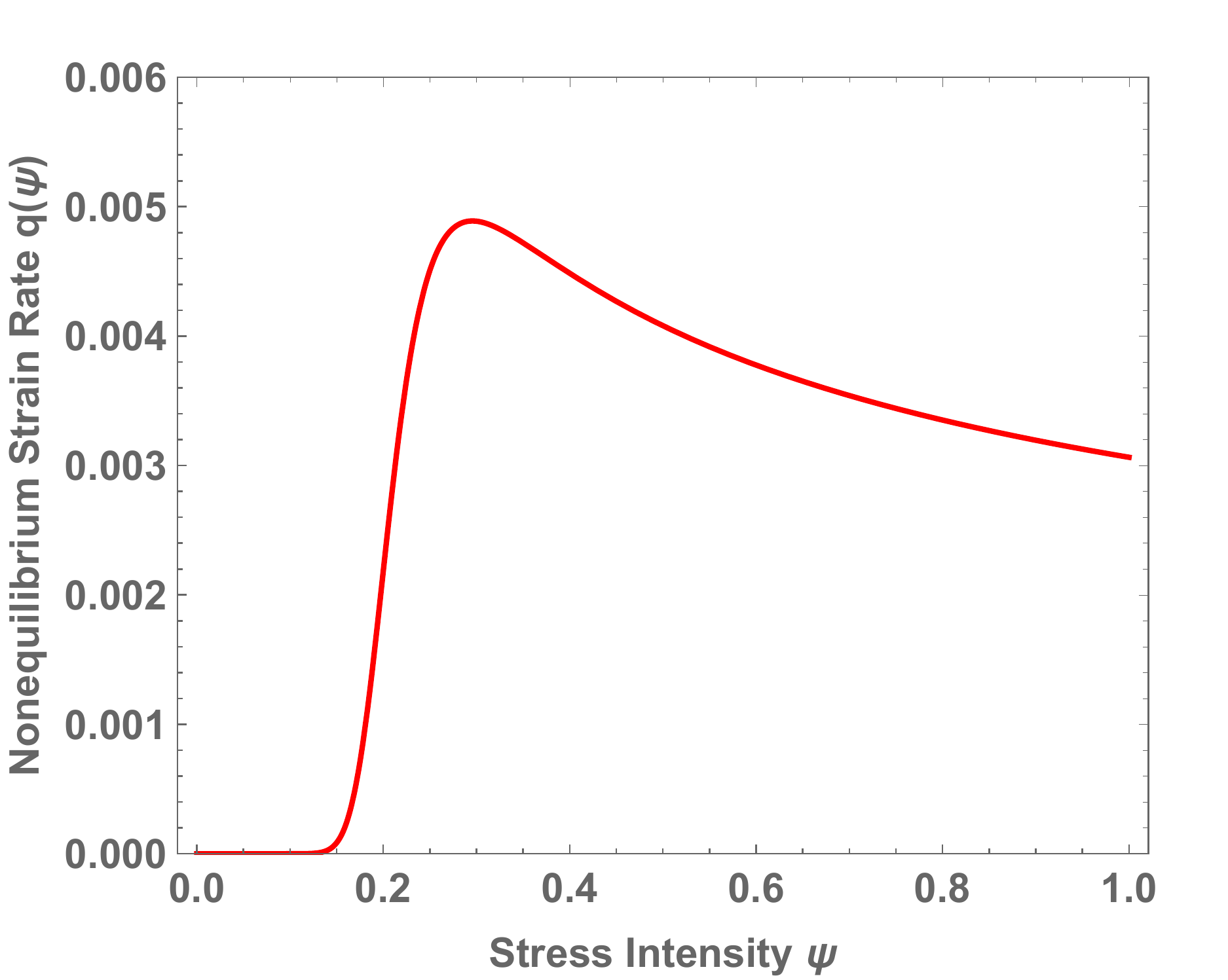}
\caption{Rate of plastic deformation $q[\tilde s_0(\psi),\tilde\rho(\psi),T(\psi)]$ as a function of $\psi$ for  $T_0 = 460\,K$.  Note that the abrupt increase in $q$ at the yield stress of about $\tilde s_y \cong 0.2$ corresponds to the onset of tip-shielding seen in Fig.\ref{BDGFig2}.}   \label{BDGFig4}
 \end{center}
\end{figure}

{\it Numerical Results.}  Figure \ref{BDGFig2} is a graph of the tip stress $\tilde s_0(\psi)$, computed using the parameters listed above, with initial temperature $T_0 = 460\,K$, and for values of $\psi$ out to about $3.34$.  This curve illustrates several important features of the theory, especially the brittle-ductile transition.  Note that the stress rises linearly from $\psi = 0$.  This is the initial elastic response predicted by Eq.(\ref{tipstress}) for $\bar\nu = 1$ (no boundary layer).  The curve starts to bend over at $\psi \cong .2$ in a relatively smooth plastic yielding transition.  That transition can be seen in more detail in Fig. \ref{BDGFig4},  where I have plotted the rate of plastic deformation $q[\tilde s_0(\psi),\tilde\rho(\psi),T(\psi)]$ as a function of $\psi$ in the range $0 < \psi < 1$.  This transition is not infinitely sharp; but it is sharp enough to justify the approximations made in deriving the expressions for the yield stress in Eqs.(\ref{sy2}) and (\ref{sy3}) and in deriving the equation of motion for $\tilde s_0$ in Eqs.(\ref{s0dot}) and (\ref{tipstress}).

\begin{figure}[h]
\begin{center}
\includegraphics[width=\linewidth] {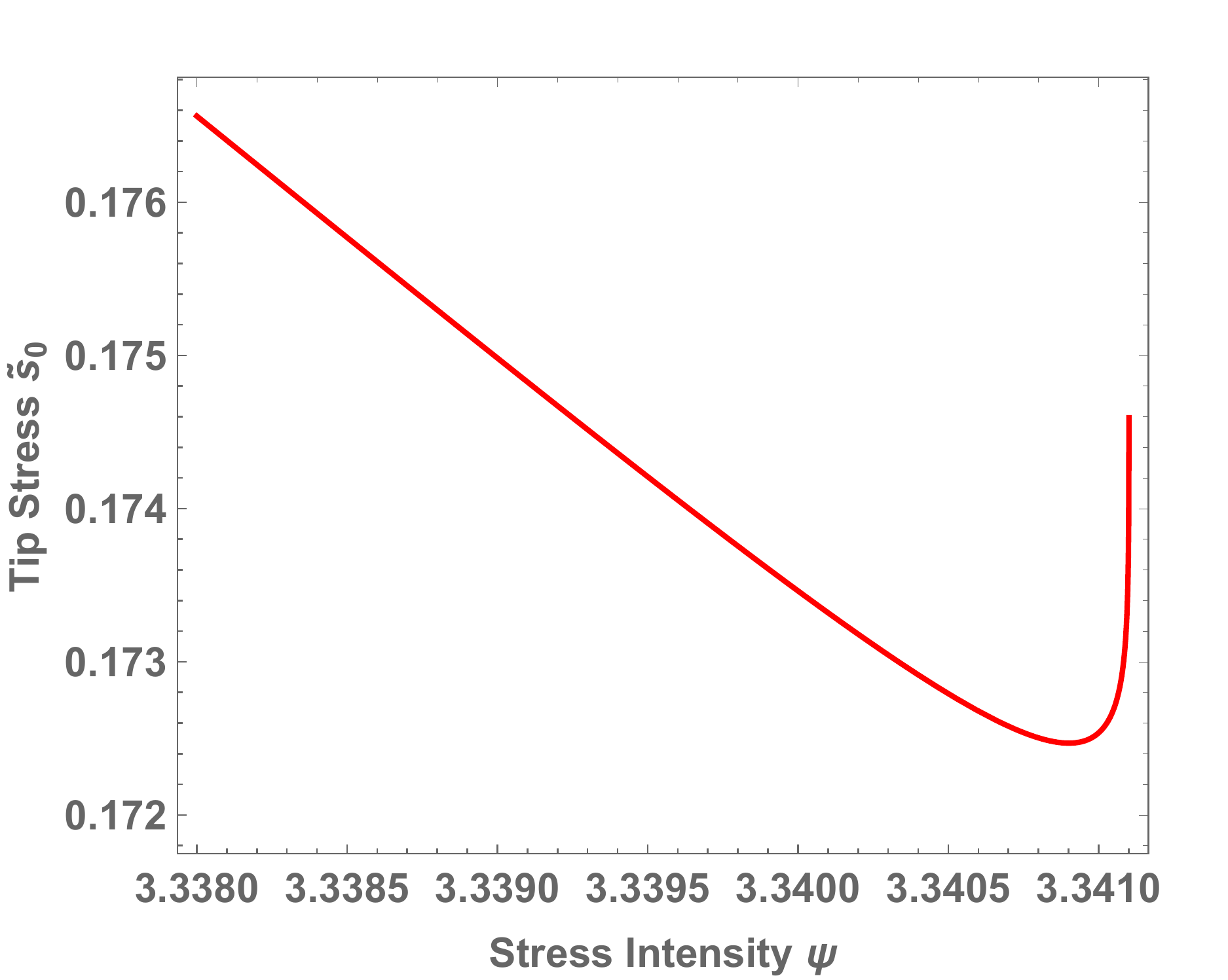}
\caption{Expanded graph of $\tilde s_0(\psi)$ at the divergence near $\psi \cong 3.34$, originally seen as a sharp discontinuity in Fig.2. }   \label{BDGFig5}
 \end{center}
\end{figure}

\begin{figure}[h]
\begin{center}
\includegraphics[width=\linewidth] {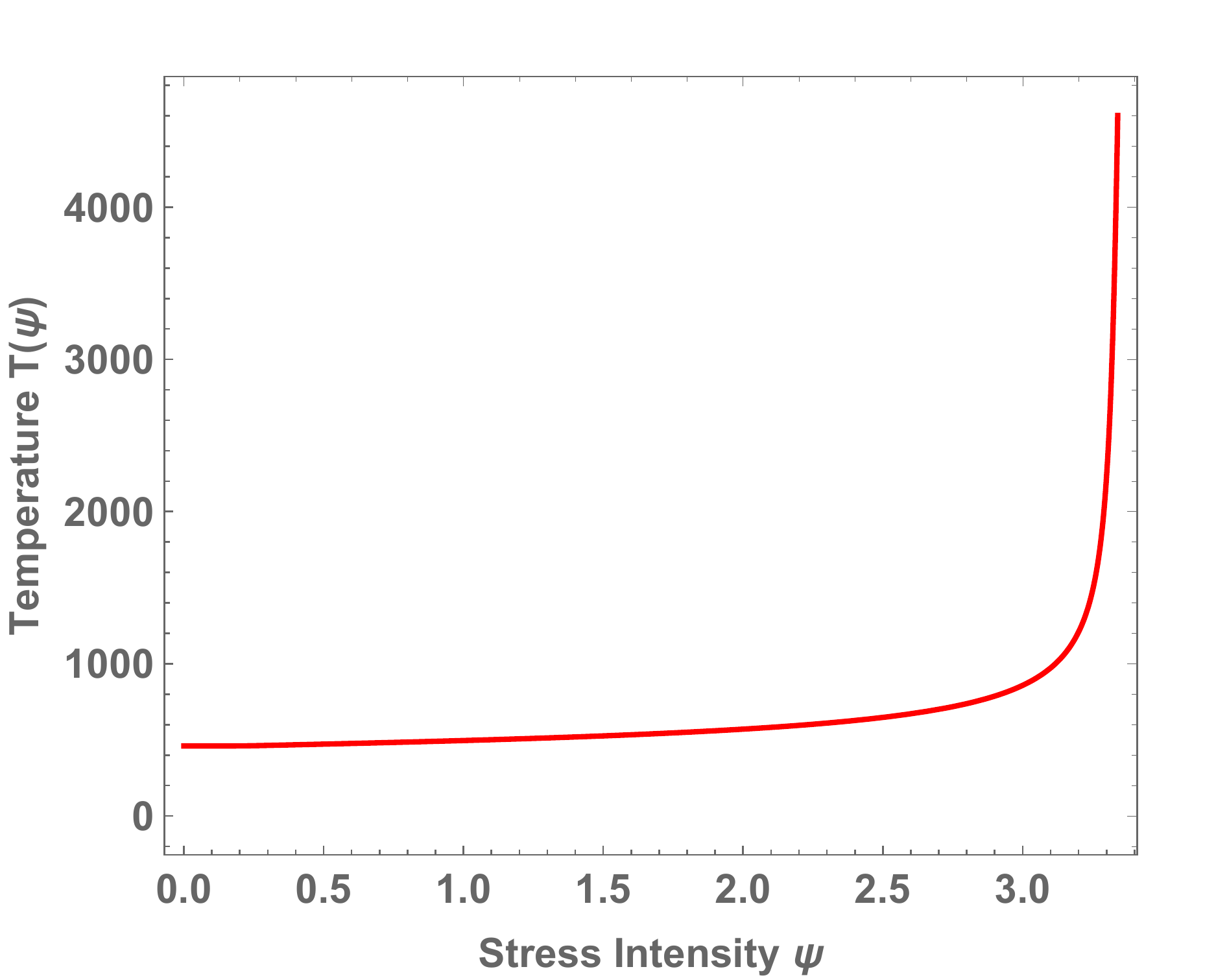}
\caption{Theoretical tip temperature $T(\psi)$ for $T_0 = 460\,K$. Note the strong divergence at the onset of the ductile instability at $\psi \cong 3.34$. }   \label{BDGFig6}
 \end{center}
\end{figure}

The second dominant feature of the stress curve in Fig.\ref{BDGFig2} is the strong singularity at $\psi \cong 3.34$.  This is where the plastic boundary layer near the notch tip breaks down and the tip stress rises abruptly, indicating the onset of large-scale ductile failure.  This behavior is driven by the divergence of the function $\Lambda(\bar\nu)$ at $\bar\nu \cong 5.1$.  An expanded picture of this divergence of $\tilde s_0(\psi)$ is shown in Fig. \ref{BDGFig5}, making it clear that this rapid upturn is mathematically smooth.  This divergence of the stress is accompanied by a divergence of the temperature $T(\psi)$ as shown in Fig.\ref{BDGFig6}.  

\begin{figure}[h]
\begin{center}
\includegraphics[width=\linewidth] {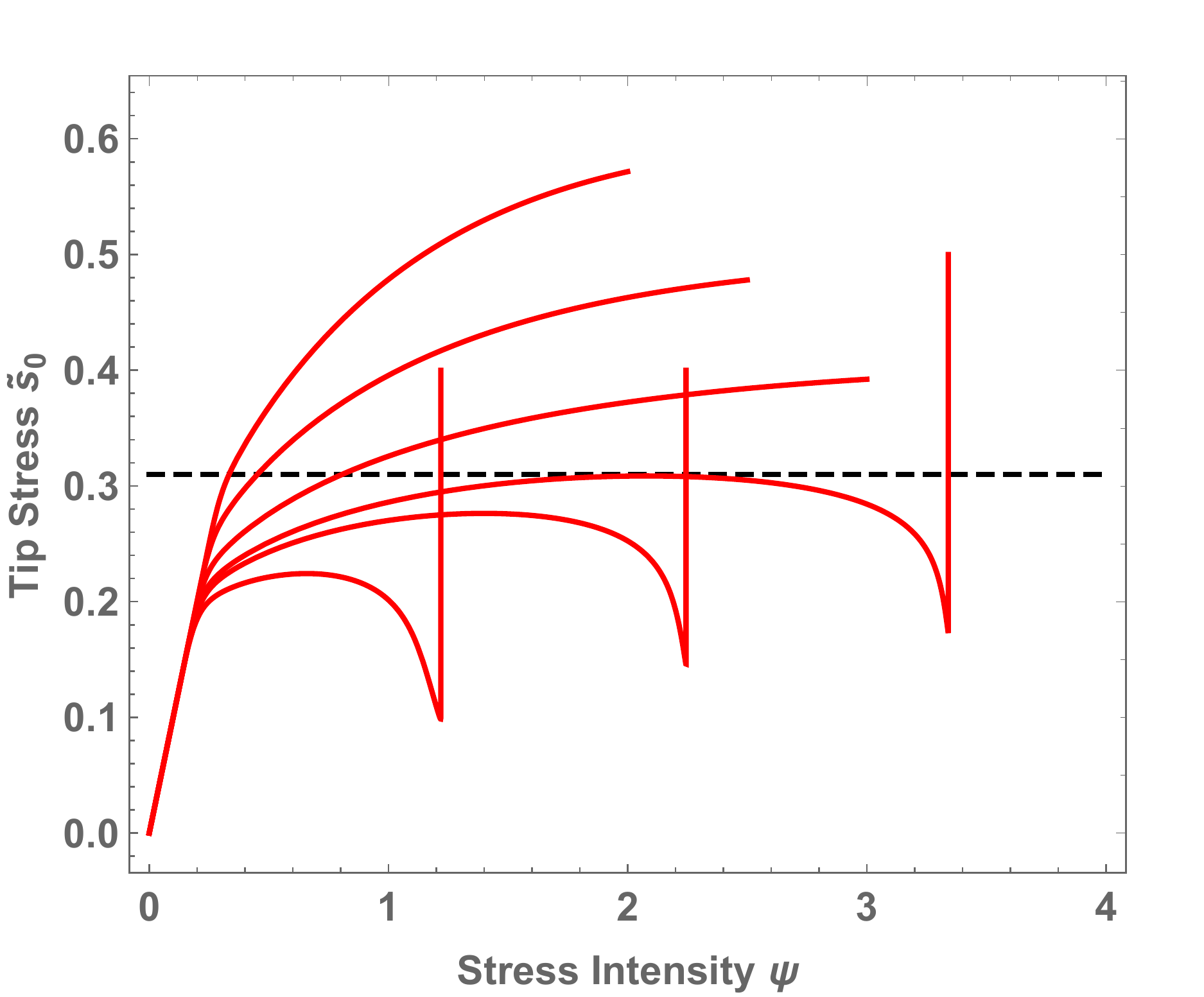}
\caption{Tip stresses $\tilde s_0(\psi)$ for temperatures $T_0 = 100\, K, 200\, K, 350\, K, 460\, K, 500\, K,$ and $600\, K$, from top to bottom, plus a dashed line at the breaking stress $\tilde s_0 = \tilde{s}_c = 0.31$. }   \label{BDGFig7}
 \end{center}
\end{figure}

The last assumption needed in order to compute the theoretical values of the fracture toughness shown in Fig.\ref{BDGFig1} is that  brittle fracture is initiated when the dimensionless deviatoric tip stress $\tilde s_0$ reaches a critical value, say $\tilde s_c$.  Figure \ref{BDGFig7} shows six different tip-stress functions $\tilde s_0(\psi)$ for initial temperatures $T_0 = 100\, K, 200\, K, 350\, K, 460\, K, 500\, K,$ and $600\, K$, and a horizontal dashed line at $\tilde s = \tilde{s}_c = 0.31$, which is my estimate for the breaking stress in this case.   The values of the fracture toughness ${\cal K}_c$ in Fig.\ref{BDGFig1}  are proportional to the values of the stress intensity $\psi_c$ at which the stress curves cross $\tilde{s}_c$, that is, $\tilde s_0(\psi_c) = \tilde{s}_c$.  From this analysis, I estimate that ${\cal K}_c \cong 20\, \psi_c$.  This proportionality factor should depend only on the instrumentation and not the sample preparation; thus I use it throughout this paper.

The theoretical fracture-toughness curve shown in Fig. \ref{BDGFig1} was constructed by computing $\tilde s_0(\psi)$ at fourteen different initial temperatures $T_0$. For clarity, only six of those curves are shown in Fig.\ref{BDGFig7}.  At the lowest temperatures, plastic shielding is negligable and the fracture toughness levels off at $\psi_c \approx \tilde{s}_c$, that is, at  ${\cal K}_c \cong 6$.  At the highest initial temperatures, the shielding effect is strong and the tip stress does not reach $\tilde{s}_c$ before the boundary layer undergoes its thermal instability indicated by the (nearly) vertical lines in the figure.  Those intersections automatically give us values of $\psi_c$ on the ductile part of the curve.   The one exception is the $\tilde s_0(\psi)$ curve at $T_0 = 460\,K$ whose peak is tangent to the horizontal line at $\tilde{s}_c = 0.31$.  Here I assume that the toughness jumps discontinuously from the tangent point to the thermal instability point, and that this is what is interpreted experimentally as the brittle-ductile transition.  Surely this is only an approximation. 

On the whole, given the uncertainties in both the theory and the experiments, this nontrivial agreement with experiment over a temperature range of $500\,K$ leads me to believe that this part of the theoretical picture is basically correct.  But the picture is less clear for the non-predeformed crystals.  

\section{Fracture Toughness of Non-Predeformed Crystals: Theory}

Gumbsch's measurements of the fracture toughness of non-predeformed crystalline tungsten at several different driving rates bring us into a qualitatively new, fundamentally uncertain, and highly interesting dynamical situation.  Initially, these notch tips are not surrounded by substantial densities of entangled dislocations.  On the contrary, with increasing opening stresses, they are first driven to emit dislocations one at a time.  As those dislocations move away from the tip, they produce plastic deformation of the material and  the tip shape. This situation immediately poses a challenge for any first-principles analysis.  

To see what is happening, start by assuming that a single dislocation in the neighborhood of the notch tip is subject to a drag force, so that its velocity (say, in the $\theta$ direction) is
\begin{equation}
\label{vdrag}
v_{\theta}^{d\!r\!a\!g}={ b\,s_{\theta,\theta}\over \eta(T)\,\tau_{d\!r\!a\!g}},
\end{equation}
where $\eta(T)$ is a $T$-dependent drag coefficient with the dimensions of stress; $\tau_{d\!r\!a\!g}$ is the associated time constant; and $s_{\theta,\theta}$ is the  $\theta,\theta$ component of the deviatoric stress introduced in Eq.(\ref{stress1}). I am greatly oversimplifying this situation.  Strictly speaking, $s_{\theta,\theta}$ should be the projection of the deviatoric stress  onto a glide plane, and I should be averaging over orientations of glide planes relative to the orientation of the notch tip.  But the approximation in Eq.(\ref{vdrag}) is already good enough to reveal an important shortcoming of the theory as it stands.  

Suppose that we were dealing with a set of noninteracting dislocations moving in a uniform environment.  Then Orowan's formula -- essentially dimensional analysis -- would tell us that the corresponding rate of plastic deformation (see Eq.(\ref{RODT})) is 
\begin{equation}
\label{Ddrag}
D^{d\!r\!a\!g}_{\theta,\theta} = -D^{d\!r\!a\!g}_{\zeta,\zeta}\equiv D^{d\!r\!a\!g} \propto \rho_D\,b\,v_{\theta}^{d\!r\!a\!g}\propto {\rho_D\,  b^2\,s_{\theta,\theta}\over \eta(T)\,\tau_{d\!r\!a\!g}},~~~~~
\end{equation}
where $\rho_D$ is the areal density of these ``dragged'' dislocations; and the ``$\propto$'' symbols imply as-yet unknown, dimensionless proportionality factors.

Now use Eq.(\ref{kappadot})  to compute the part of the equation of motion for the curvature $\kappa$ arising from this deformation rate, and transform to the dimensionless variables introduced in Eqs.(\ref{psidef}) and (\ref{dotpsidef}).  I find that
\begin{equation}
\label{kappadotdrag}
{1\over \kappa^{3/2}}\Bigl({d\kappa\over d\psi}\Bigr)_{d\!r\!a\!g} \propto {\xi\,{\tilde\rho}_D\,\psi\over \tilde{\eta}(T)},
\end{equation}
where ${\tilde\rho}_D = a^2 \rho_D$.  The dimensionless drag coefficient is
\begin{equation}
{1\over \tilde{\eta}(T)} = {\mu_T\over \eta(T)}\,\Bigl({b\over a}\Bigr)^2 {\tau_{pl}\over \tau_{d\!r\!a\!g}}.
\end{equation}
Also, in this limit with no plastic shielding, we know that the tip stress is ${\tilde s}_0 = \psi\,\sqrt{\kappa}$, so that 
\begin{equation}
\label{sdotdrag}
\Bigl({d{\tilde s}_0\over d\psi}\Bigr)_ {d\!r\!a\!g}\propto {\xi\,{\tilde\rho}_D\,\psi^2\,\kappa\over \tilde{\eta}(T)}.
\end{equation}

Equations (\ref{kappadotdrag}) and (\ref{sdotdrag}) have exactly the {\it wrong}  dependence on the driving rate.  Remember that $\xi \equiv \tau_{ex}/\tau_{pl}$, so that the external driving rate is proportional to $1/\xi$. These equations seem to tell us that the approach to fracture becomes slower with increasing driving rate; but experiment and common sense tell us the opposite.   

The problem here is that Orowan's simple dimensional analysis is unlikely to be correct in a situation where there are multiple competing length and time scales.  The separation between the dislocations may be substantially larger than the radius of curvature of the notch tip. Moreover, these dislocations may be escaping from the tip region in times shorter than the time scale for tip growth. This is far too complex a situation to be modeled by the mathematical approximations used here.  In order to make progress, I must make  some simplifying, {\it ad hoc} assumptions.

First, I propose including in the Orowan relation for the drag term an extra dimensionless factor $1/\xi^2$, that is, rewriting Eq.(\ref{Ddrag})  in the form
\begin{equation}
\label{Ddrag2}
D^{d\!r\!a\!g} =  {\rho_D\,  b^2\,s_{\theta,\theta}\over \xi^2\,\tilde{\eta}(T)\,\tau_{d\!r\!a\!g}},
\end{equation}
where I have absorbed some other dimensionless factor into $\tilde{\eta}(T)$. Then Eqs.(\ref{kappadotdrag}) and (\ref{sdotdrag}) become 
\begin{equation}
\label{kappadotdrag2}
{1\over \kappa^{3/2}}\Bigl({d\kappa\over d\psi}\Bigr)_{d\!r\!a\!g} = {{\tilde\rho}_D\,\psi\over \xi\,\tilde{\eta}(T)}
\end{equation}
and
\begin{equation}
\label{sdotdrag2}
\Bigl({d{\tilde s}_0\over d\psi}\Bigr)_ {d\!r\!a\!g}= {{\tilde\rho}_D\,\psi^2\,\kappa\over \xi\,\tilde{\eta}(T)}.
\end{equation}
In effect, I am assuming that the generalized Orowan formula for dilute dislocations near a sharp notch tip contains, not the number of dislocations in a square of side $b$, but the number in a square whose side length is proportional to the driving rate. In other words, as the system is driven faster, more of the dislocations near the tip contribute to the plastic deformation.

Part of my rationale here is that, with this $\xi^2$ correction, the entire temperature and driving-rate dependence of this part of the fracture-toughness calculation resides in the product $\xi\,\tilde{\eta}(T)$.  Thus, if $\tilde{\eta}(T)$ is a thermal activation factor of the form $\exp(- T_D/T)$, then we automatically obtain Gumbsch's scaling law relating temperature, driving rate, and fracture toughness throughout the low-temperature, brittle regime of these measurements.

My second proposed modification is to assume that there are effectively two distinct populations of dislocations described by dimensionless densities $\tilde\rho$ and $\tilde\rho_D$ (as before, in units $a^{-2}$).  The ``late-stage'' density $\tilde\rho$ is the same as $\tilde\rho$ in the preceding parts of this paper.  It describes the dislocations that become entangled with each other and shield the notch tip during the later stages of the brittle-ductile transitions.  The ``early-stage'' density $\tilde\rho_D$ is the same as $\rho_D/a^2$ in Eq.(\ref{Ddrag}).  I assume that these two families of dislocations can be dealt with separately from each other.  

In particular, I assume that the late-stage disentanglement dynamics is not modified by the early-stage drag stresses in the way that I found was important in my analysis of the Livermore molecular dynamics simulations.\cite{BULATOV-17,JSL-18}  There, in a spatially uniform situation, I found an early-stage regime in which the drag forces significantly modified the times between pinning and depinning events, and thus measureably affected the plastic deformation rate.  I shall ignore that effect here, and simply add the  $\tilde\rho_D$ terms given by Eqs.(\ref{kappadotdrag2}) and (\ref{sdotdrag2}) to the right-hand sides of Eqs.(\ref {kappadot3}) and (\ref{tipstress}).

The augmented versions of the latter equations now become:
\begin{equation}
\label{kappadot4}
{1\over \kappa^{3/2}} {d\kappa\over d\psi}= c_0\Bigl[{(\bar\nu -1)^2\over 3}{\Lambda(\bar\nu)\over\bar\nu^3} + \Bigl({2 \bar\nu -1\over \bar\nu^2}\Bigr) \Bigr]+{{\tilde\rho}_D\,\psi\over \xi\,\tilde{\eta}(T)},~~~~~~
\end{equation}
and
\begin{equation}
\label{tipstress2}
{d \tilde s_0\over d\psi} = - {\xi\over c_0} q(\tilde s_0,\tilde\rho,T) + \sqrt{\kappa}\, {\Lambda(\bar\nu)\over\bar\nu^3}+{{\tilde\rho}_D\,\psi^2\,\kappa\over \xi\,\tilde{\eta}(T)}.
\end{equation}

The equation of motion for the dislocation density $\tilde\rho$, Eq.(\ref{rho-psi}), remains unchanged except that the energy-conversion prefactor $A$ is explicitly written here as a function of $\xi$ to include corrections similar to those introduced in the equation of motion for $\tilde{\rho}_D$.  
\begin{equation}
\label{rho-psi2}
{d \tilde\rho\over d\psi} = A(\xi)\,\xi \,q(\tilde s_0,\tilde\rho,T)\,\tilde s_0(\psi)\,\Bigl[1 - {\tilde\rho(\psi)\over \tilde\rho_{\infty}}\Bigr].
\end{equation}

My proposed equation of motion for ${\tilde\rho}_D$ is:
\begin{equation}
\label{rhoD-psi}
{d{\tilde\rho}_D\over d\psi} = A_D\,\,{\psi^2\,\tilde{\rho}_D(\psi)\,\kappa(\psi)\over \xi\,\tilde{\eta}(T)}\,\Bigl[1-{{\tilde\rho}_D(\psi)\over {\tilde\rho}_c}\Bigr].
\end{equation}
This analog of Eq.(\ref{rho-psi2}) says that the rate at which dislocations are created is proportional to the energy flow to the tip estimated using Eq.(\ref{Ddrag2}). But here, instead of the last factor in Eq.(\ref{rho-psi2}) which assures approach to steady-state TDT equilibrium, I have multiplied this formation rate by $[1-{\tilde\rho}_D(\psi)/{\tilde\rho}_c]$, with a small value of ${\tilde\rho}_c$, to account for the fact that most of these early-stage dislocations escape from the tip region.

The equation of motion for the temperature $T$, Eq.(\ref{TQ}), remains as is, with the understanding that the prefactor $C(T)$ will have to be adjusted to fit different temperature regimes, just as in the theories of thermal softening and adiabatic shear banding cited previously.\cite{LTL-17,LTL-18} In principle, there should be no $\xi$-dependent correction here because there is no analog of the Orowan factor for normal thermal fluctuations; the thermal length and time scales are always microscopic. But different values of $\xi$ imply different dynamic and thermal ranges, and thus somewhat different values of $C(T)$.

\begin{figure}[h]
\begin{center}
\includegraphics[width=\linewidth] {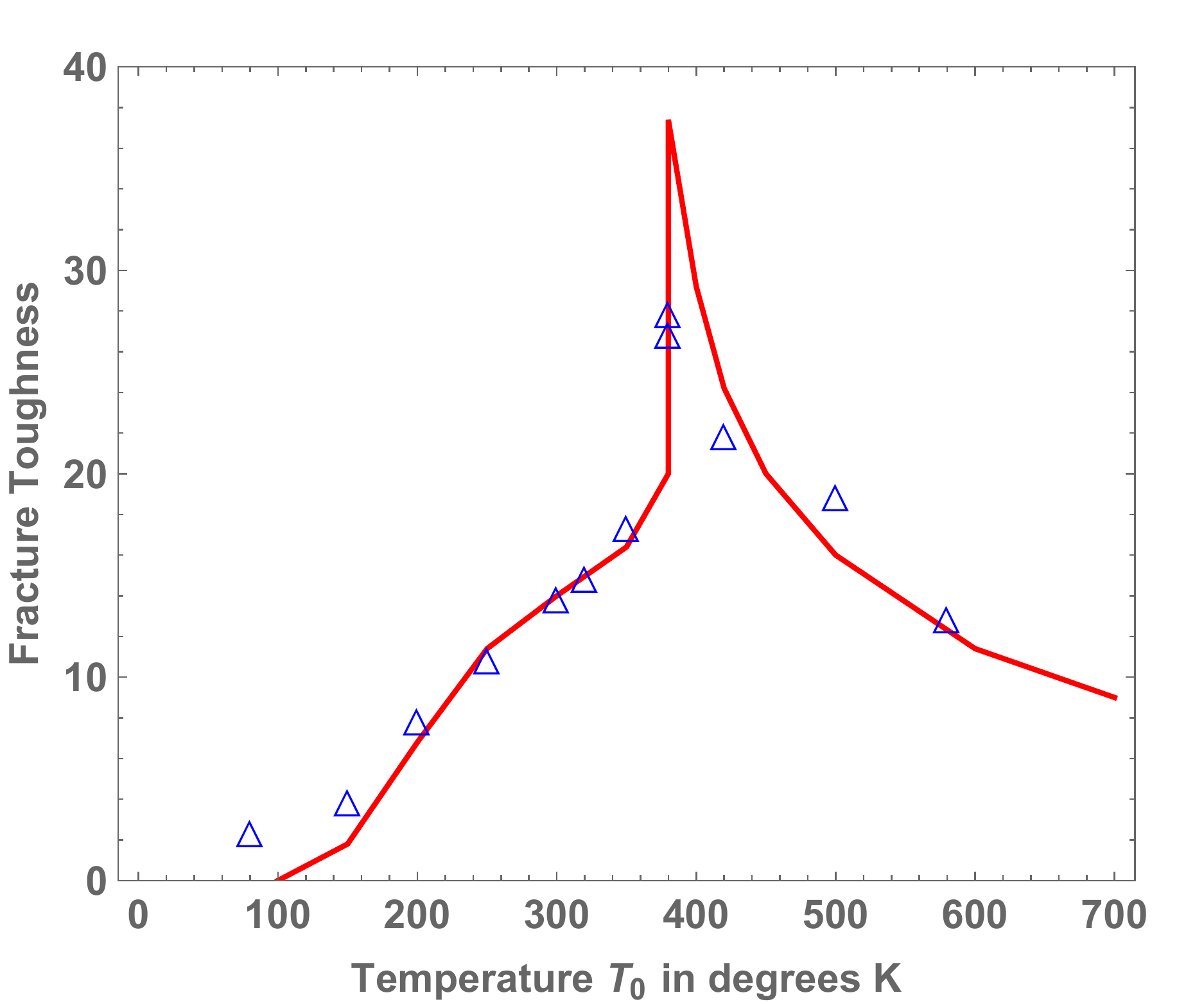}
\caption{Experimental data for fracture toughness as a function of temperature (open triangles), and theoretical prediction (solid curve), for non-predeformed crystalline tungsten. The fracture toughness  ${\cal K}_c$ is in units $M\!Pa\,\,m^{1/2}$. The loading rate, $0.1\, M\!Pa\,\,m^{1/2}\,s^{-1}$, {\it i.e.}  $\xi = 1$, is the same as in Fig. 1 for the predeformed case.}   \label{BDGFig8}
 \end{center}
\end{figure}

\section{Fracture Toughness of Non-Predeformed Crystals: Comparison Between Theory and Experiment}

\begin{figure}[h]
\begin{center}
\includegraphics[width=\linewidth] {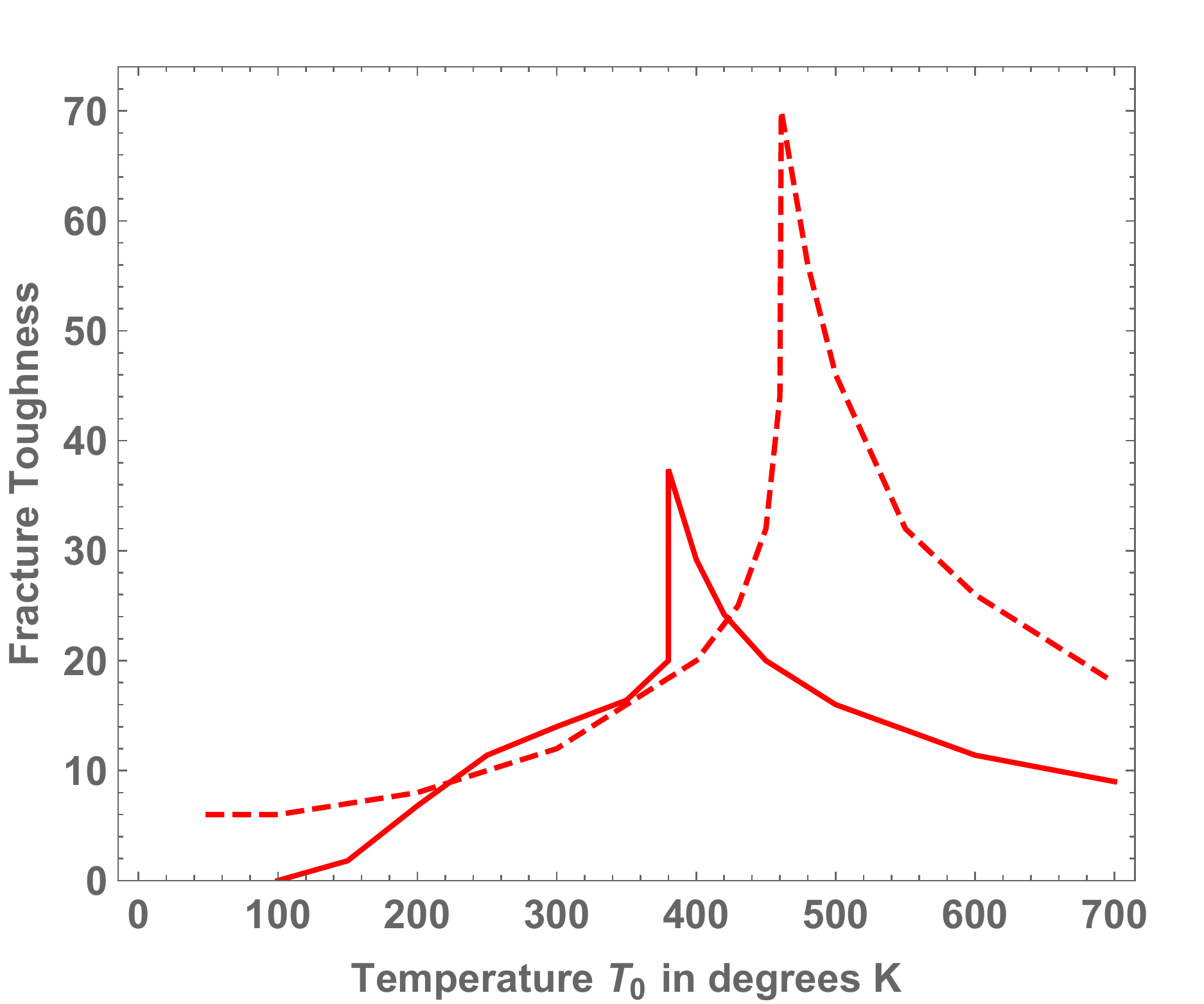}
\caption{Comparison between theoretical fracture-toughness curves for the predeformed case (dashed curve) and the non-predeformed case (solid curve), both at the same loading rate $\xi = 1$. }   \label{BDGFig9}
 \end{center}
\end{figure}

Turn now to Gumbsch's data for non-predeformed (i.e. non-prehardened) tungsten crystals, specifically as shown in Fig.4 of Ref. \cite{GUMBSCH-03} for loading rates $0.1,\, 0.4,$ and $1.0\,M\!Pa\,\,m^{1/2}\,s^{-1}$ ($\xi = 1.0,\,0.25$, and $0.10$).    Figure \ref{BDGFig8} shows my theoretical fit to Gumbsch's data; and Fig. \ref{BDGFig9} is a theoretical comparison of the two cases -- predeformed and non-predeformed -- both for $\xi = 1$, comparable to the experimental comparison shown in Gumbsch's Fig.2.  

{\it System Parameters}:  Parameters that remain unchanged (for all $\xi$) from the prehardened analysis are:  $T_P = 36,000\,K$; $\tilde\rho_{\infty} = e^{-4}$; $\tilde\rho_{min} = 0.1\, \tilde\rho_{\infty}$; $c_0 \equiv \mu_T/2\,\mu = 0.01$; and the ratio of the measured fracture toughness to the dimensionless critical stress intensity, ${\cal K}_c/\psi_c = 20$.  
 
 In the thermal conversion factor in Eq.(\ref{TQ2}), I have chosen to keep $T_A= 3500\,K$ in all cases, but allow moderate changes in the prefactor $C_0$.  For the non-prehardened case shown in Fig. \ref{BDGFig8}, $C_0 = 2\,\times\,10^8$ (somewhat larger than $C_0 = 0.7\,\times\,10^8$ for the prehardened case).  
 
For all cases (including all $\xi$), my choice of the drag coefficient is
\begin{equation}
\tilde{\eta}(T) = 1.2\,\,e^{-T_D/T},
\end{equation}
with $T_D = 2200\,K$ (Gumbsch's scaling temperature).  For simplicity, I have chosen not to add a temperature-independent phenomenological constant to this formula; thus the theoretical fracture toughness (e.g. in Fig.\ref{BDGFig8}) vanishes too rapidly as $T_0\to 0$.  

An important parameter for the non-prehardened cases is the limiting value of the density of the early-stage dislocations.  In Eq.(\ref{rhoD-psi}),  ${\tilde\rho}_c = 0.01\,\tilde\rho_{\infty}$ seems to work well for all $\xi$, implying a high escape probability.  However, initial values for both $\tilde\rho(\psi)$ and $\tilde{\rho}_D(\psi)$ must be effectively zero for non-deformed crystals.   I have taken them to be equal to $10^{-7}$ (negligably small).  

These choices of initial conditions raise an issue about the $\xi$ dependence of the conversion factor $A(\xi)$ in the equation of motion for $\tilde{\rho}(\psi)$, Eq.(\ref{rho-psi2}).  For computing the toughness curve in Fig.\ref{BDGFig8}, I have had to use $A(1)=100$ (in contrast to $A = 10$ for the prehardened case).  Here I am pushing my {\it ad hoc} model to its limits in order to start from  $\tilde{\rho}(0) \approx 0$ at $\psi = 0$ and generate large enough values of  $\tilde{\rho}(\psi)$ to shield the tip at $\psi \approx 1$. This problem is not so severe at the higher driving rates.

\begin{figure}[h]
\begin{center}
\includegraphics[width=\linewidth] {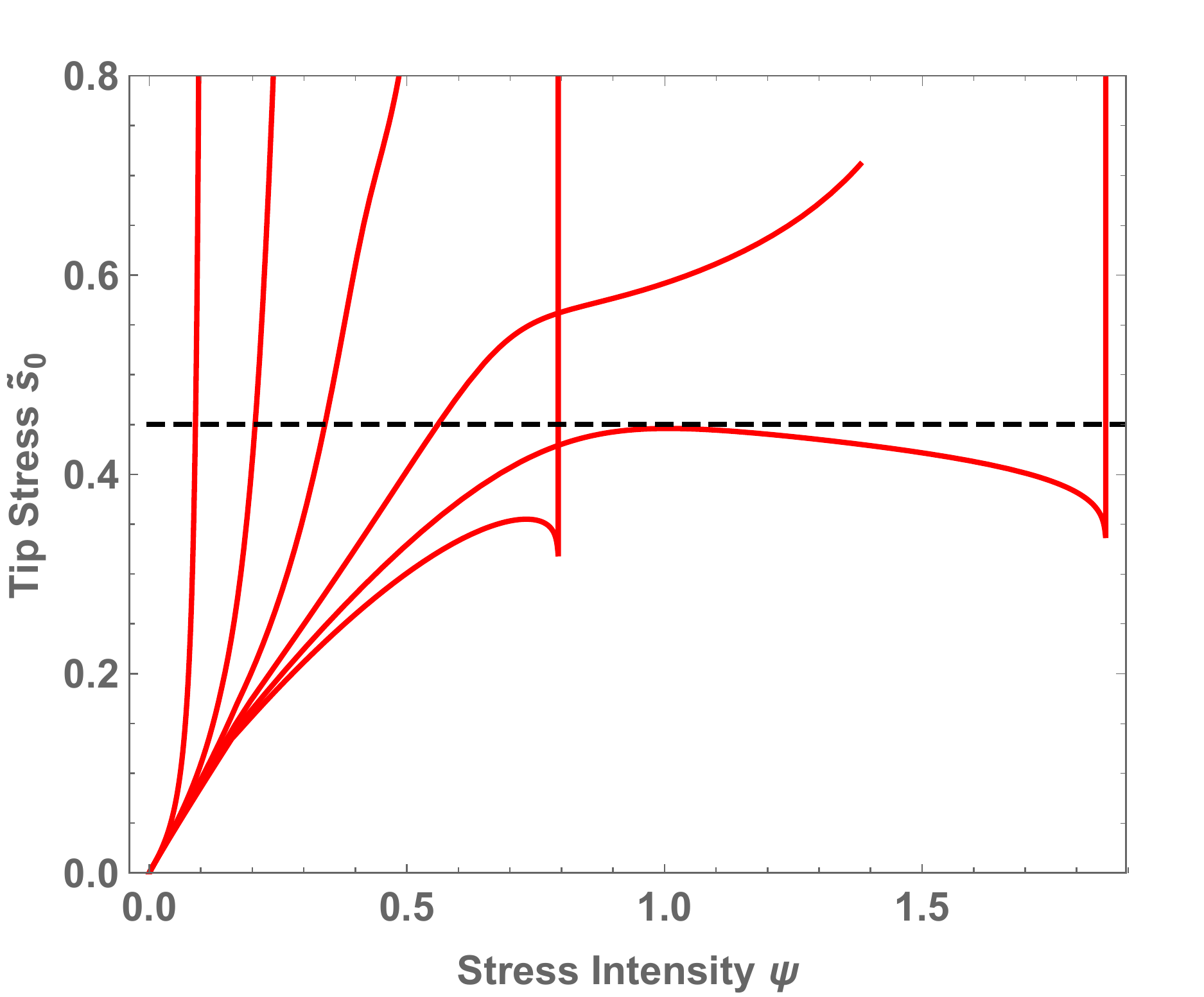}
\caption{Tip stresses $\tilde s_0(\psi)$ for temperatures $T_0 = 150\, K, 175\,K, 200\, K, 250\,K,\, 380\,K $ and $500\, K$, from top left to bottom right, plus a dashed line at the breaking stress $\tilde s_0 = \tilde{s}_c = 0.45$. }   \label{BDGFig10}
 \end{center}
\end{figure}
\begin{figure}[h]
\begin{center}
\includegraphics[width=\linewidth] {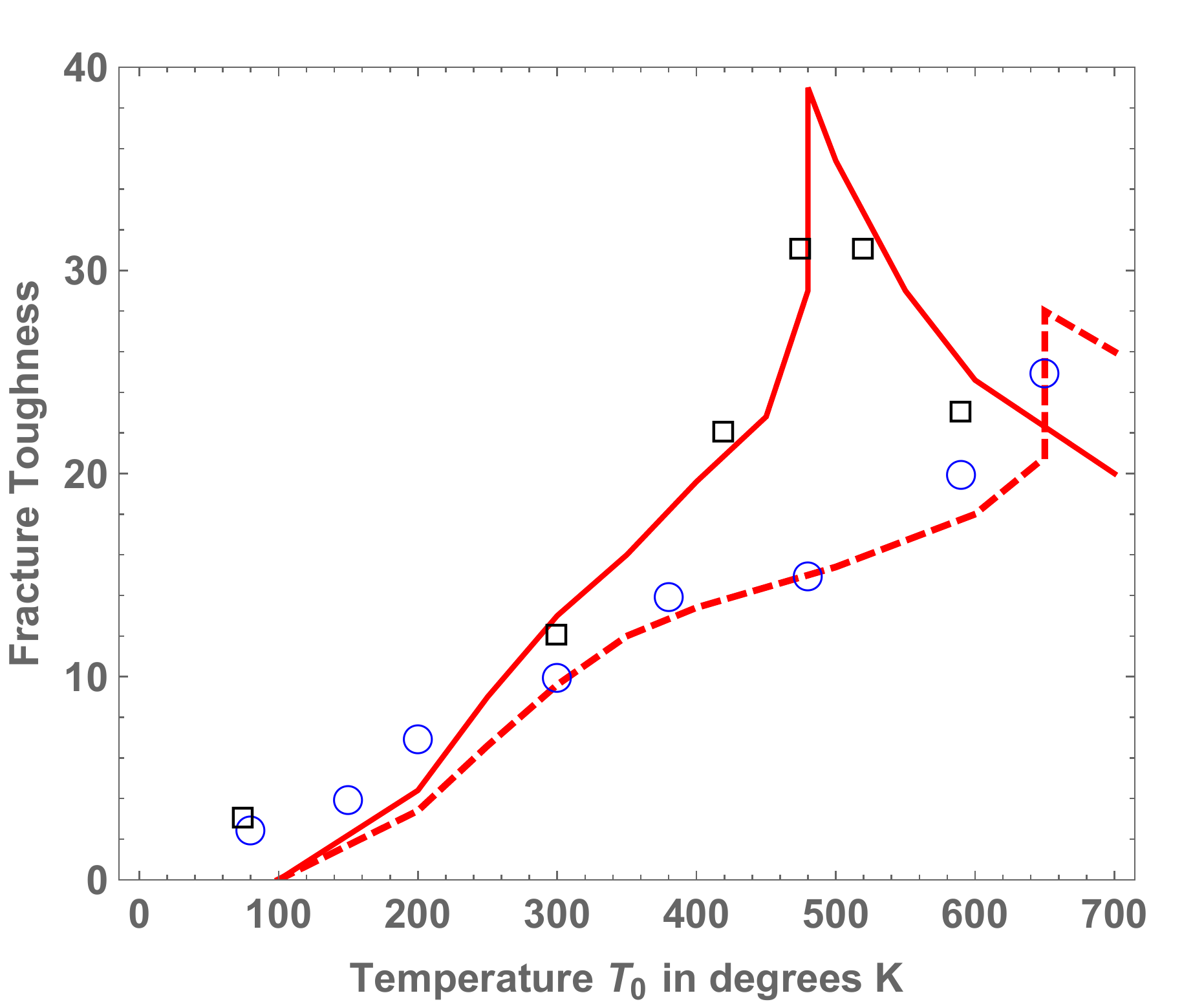}
\caption{Experimental data for fracture toughnesses as functions of temperature (open squares and open circles), and theoretical predictions (solid and dashed curves) for non-predeformed crystalline tungsten at loading rates  $0.4$ and $1.0\, M\!Pa\,\,m^{1/2}\,s^{-1}$ , {\it i.e.} $\xi = 0.25$ and $0.1$ respectively.}   \label{BDGFig11}
 \end{center}
\end{figure}

{\it Numerical Results.}  Now, look at Figs. \ref{BDGFig8} - \ref{BDGFig10} and focus on the main features of the non-predeformed case at the smallest driving rate, $\xi = 1$. The comparison seen in Fig.\ref{BDGFig9} (also in Gumbsch \cite{GUMBSCH-03}, Fig.2, for the experimental points) shows that the predeformed -- i.e. ``hardened'' -- material is indeed tougher on average than the non-predeformed crystal.  But the non-predeformed crystal is tougher for a range of initial temperatures $T_0$ near its brittle-ductile transition.  A related point is that the fracture stress $\tilde{s}_c$ is somewhat larger for the non-predeformed crystals ($\tilde{s}_c \sim 0.45)$ than for the predeformed one ($\tilde{s}_c \sim 0.3)$.  Apparently, the work-hardening process introduces new defects, or weakens existing ones, thus reducing $\tilde{s}_c$.  One of my considerations in fitting the fracture-toughness data for the non-predeformed crystals at higher driving rates is that $\tilde{s}_c$ must be roughly the same for all values of $\xi$.

The tip-stress functions $\tilde{s}_0(\psi)$ shown in Fig.\ref{BDGFig10} for the non-predeformed crystals at $\xi = 1$ are analogous to, but interestingly different from those for the predeformed case in Fig. \ref{BDGFig7}.  Most obviously, the stress curves in Fig.\ref{BDGFig10} for the lower temperatures show the tip-sharpening instabilities predicted by Eqs.(\ref{kappadotdrag2})  and (\ref{sdotdrag2}).  These equations can be solved analytically when $\tilde{\rho}_D \to \tilde{\rho}_c$ is a constant, predicting the observed divergence of the stress and unbounded tip sharpening -- limited only by the breaking stress $\tilde{s}_c$ as shown in the figure.  With increasing $T_0$, the factor $\tilde{\eta}^{-1}$ decreases rapidly, and the sharpening effect is supressed.  The result is the pronounced shoulder on the low-temperature side of the fracture-toughness curve in Fig. \ref{BDGFig8}.  The behavior at larger values of $T_0$, up to and beyond the brittle-ductile transition at $T_0 \cong 380\,K$, is qualitively the same as that for the predeformed situation because, by that stage in the process, the higher-density, late-stage dislocations are dominant. 

Figure \ref{BDGFig11} shows the fracture-toughness curves, both experimental and theoretical, for the non-predeformed tungsten crystals loaded rapidly at $\xi = 0.25$ and $0.10$.  Parameter adjustments for these cases are relatively modest.  For the dislocation-creation coefficients $A(\xi)$ in Eq.(\ref{rho-psi2}), $A(0.25) = 40$ and $A(0.1) = 100$, which means that $\xi\,A(\xi) \cong 10$ for both of these cases as well as for the predeformed one with $\xi = 1$.  The thermal prefactors defined in Eq.(\ref{TQ2}) are $C_0 = 0.6 \,\times\,10^8$ for $\xi = 0.25$ and $C_0 =0.3\,\times\,10^8$ for $\xi = 0.1$. 
For these two values of $\xi$, the breaking stresses are $\tilde{s}_c = 0.45$ and $0.49$ respectively. 

In view of the various uncertainties, including the sparse experimental data sets near the brittle-ductile transitions, the agreement between theory and experiment in Fig.\ref{BDGFig11} seems to be quite satisfactory.  Remember that it is these experiments, together with the data for $\xi = 1$ shown in Fig. \ref{BDGFig8}, that support Gumbsch's scaling law as shown in Fig.5 of Ref. \cite{GUMBSCH-03}.  

\section{Discussion}

The theory presented here raises many questions.  Most immediately: Is there a first-principles derivation of my proposed modification of the Orowan relation in Eq.(\ref{Ddrag2})?  What is its range of validity?  Is there any way, short of full-scale, position dependent molecular-dynamics simulations, to compute notch-tip dynamics directly in the limit of small dislocation densities?  Similarly, how do we understand the geometry-dependent modification of the energy-conversion coefficient $A(\xi)$  that determines the density of late-stage dislocations in Eq.(\ref{rho-psi2})? We know that there is not really a fundamental distinction between early-stage and late-stage dislocations.  How can we sharpen that theoretical concept?  

Regarding Eq.(\ref{Ddrag2}), remember the distinction between the present crystalline materials and the glassy materials discussed in Ref.\cite{JSLBMG-20}.  In the glassy case, irreversible plastic deformation is accomplished by reorientations of microscopic, ephemeral shear-transformation zones (STZ's).   The length and time scales associated with the latter mechanism are generally very much smaller than those associated with the dynamics of crystalline notch tips.  Thus, there is no analog of Eq.(\ref{Ddrag2}) in the glassy situation.  

Even within the approximations questioned above, we can go on to recognize  apparently successful physics-based predictions of the theory and point to other directions that might usefully be explored.  

For example, Gumbsch notes that his toughness curve for the predeformed system levels off at a fixed value at low temperatures, and that the curves for the non-predeformed cases fall below that value.  That behavior appears here in Fig. \ref{BDGFig9}.  The explanation can be seen in Fig. \ref{BDGFig7}, where the predeformed tip-stress curves rise linearly (elastically) at small $T_0$ and reach the breaking stress at $\tilde{s}_0 \cong \tilde{s}_c$ independently of $T_0$. In contrast, for the non-predeformed system, the low-temperature tip-stress curves in Fig. \ref{BDGFig10}  show clearly the predicted divergence that drives the toughness to smaller values.  

Another example (already mentioned): The theory explains the shoulders on the low-temperature sides of the non-predeformed toughness curves as  transitions between supposed early-stage and late-stage dislocation behaviors.  Can this be made more precise?

A prediction that should be tested experimentally is that the onset of ductile failure is accompanied by a sharp rise in the temperature, as seen in Fig.\ref{BDGFig7}.  

One concluding remark:  If the analysis presented here is even approximately correct, then I have answered my opening question.  The thermodynamic dislocation theory predicts the yield stress given in Eq.(\ref{sy3}), which has been confirmed experimentally. This stress generally increases as the temperature decreases.  In contrast, we see in Figs.\ref{BDGFig1}, \ref{BDGFig8}, and \ref{BDGFig11} that the fracture toughness generally decreases with decreasing, sufficiently low temperatures.  My proposed reason for this behavior is that fracture toughness is determined by the dynamic stability of notch tips -- a concept that was missing in the earlier literature.  
 
\appendix
\section{Elliptical Formulas}

This Appendix lists mathematical formulas used in the preceding analyses.  The elliptical coordinates are defined in Eq.(\ref{zetaW}).  

First, there are expressions for the diagonal elements of the rate-of-deformation tensor $D$ in terms of the elliptical material velocity components $v_{\zeta}$ and $v_{\theta}$. These are derived from more general formulas in Malvern \cite{MALVERN}.
\begin{equation}
\label{Dzeta1}
D_{\zeta\zeta} = {1\over WN}\,\left[{\partial v_{\zeta}\over \partial\zeta}+{v_{\theta}\over\zeta}\,{1\over N}\, {\partial N\over\partial\theta}\right];
\end{equation}
\begin{equation}
\label{Dtheta1}
D_{\theta\theta}={1\over WN\zeta}\,\left[{\partial v_{\theta}\over \partial\theta}+{v_{\zeta}\over N}\, {\partial \over\partial\zeta}(\zeta N)\right];
\end{equation}
where the metric function is
\begin{equation}
\label{Nmetric1}
N^2(\zeta,\theta)= 1+{m^2\over\zeta^4}-{2m\over\zeta^2}\,\cos 2\theta.
\end{equation} 
For small $\theta/\epsilon$, and $\zeta = 1$, 
\begin{equation}
\label{Nmetric2}
N \approx 2\,\epsilon\,(1 + {\theta^2\over 2\epsilon^2})
\end{equation}
Then Eqs.(\ref{Dzeta1}) and (\ref{Dtheta1})   become
\begin{equation}
\label{Dzeta2}
D_{\zeta \zeta}(\tilde x,\theta)\approx {1\over 2\epsilon\,W}\,\Bigl[{\partial v_{\zeta}\over \partial \zeta} + \Bigl({\partial v_{\theta}\over \partial\theta}\Bigr)_0\,{\theta^2\over\epsilon^2}\Bigr]\Bigl(1-{\theta^2\over 2\epsilon^2}\Bigr).
\end{equation}
and 
\begin{equation}
\label{Dtheta2}
D_{\theta\theta}
\approx {1\over 2 \epsilon W}\,\Bigl[{\partial v_{\theta}\over \partial\theta}+ \Bigl({v_{\zeta}\over\epsilon}\Bigr)_0\,\Bigl(1- {\theta^2\over \epsilon^2}\Bigr)\,\Bigr]\Bigl(1 - {\theta^2\over 2\,\epsilon^2}\Bigr).
\end{equation}\\
The notation  $(...)_0$ means that the quantity in parentheses is evaluated at $\theta = 0$. 

Next, there are the formulas for incompressible, two-dimensional elasticity that I have derived from Mushkelishvili \cite{MUSK-63}.  The following formulas assume vanishing normal stress on the surface of the elliptical hole, that is, at $\zeta = 1$.  The stress tensor $\sigma$ is given by
\begin{equation}
\label{sigmaeqn1}
\sigma_{\zeta\zeta}+\sigma_{\theta\theta}=\sigma_{\infty}\, Re\left[1+ {2(1+m)\,e^{-2i\theta}\over \zeta^2-m\,e^{-2i\theta}}\right];
\end{equation}
and
\begin{eqnarray}
\label{sigmaeqn2}
\nonumber
&&{\cal S}(\zeta,\theta)\equiv \sigma_{\theta\theta}-\sigma_{\zeta\zeta}+2i\sigma_{\zeta\theta}~~~~ \cr\\&&=\nonumber{\sigma_{\infty}\zeta^2 e^{2i\theta}\over \left(\zeta^2-m\,e^{2i\theta}\right)}\cr\\&& \times\left[1-{e^{-2i\theta}\over m\zeta^2}+{(1+m)\,e^{-2i\theta}\over \left(\zeta^2-m\,e^{-2i\theta}\right)^2} \,M(\zeta,\theta)\right]
\end{eqnarray}
where
\begin{eqnarray}
\label{Mtheta}
\nonumber
M(\zeta,\theta)&&={\zeta^2\over m}\left(1-2\,m\,e^{-2i\theta}+m^2\right)\cr\\&& + e^{-2i\theta}\left(1-2\,m\,e^{2i\theta}+m^2\right).
\end{eqnarray}
According to (\ref{sigmaeqn2}) the deviatoric stress has components
\begin{equation}
\label{devstrss}
s_{\theta\theta} =-s_{\zeta\zeta}={1\over 2}\, Re\,{\cal S}(\zeta,\theta);~~~~s_{\zeta\theta}={1\over 2}\, Im\,{\cal S}(\zeta,\theta).
\end{equation}

\begin{acknowledgments}

JSL was supported in part by the U.S. Department of Energy, Office of Basic Energy Sciences, Materials Science and Engineering Division, DE-AC05-00OR-22725, through a subcontract from Oak Ridge National Laboratory.   

\end{acknowledgments}

\end{document}